# The Scientometrics of a Triple Helix of University-Industry-Government Relations

**(Introduction to the topical issue)**

Loet Leydesdorff [1] & Martin Meyer [2]

**Abstract**

We distinguish between an internal differentiation of science and technology that focuses on instrumentalities and an external differentiation in terms of the relations of the knowledge production process to other social domains, notably governance and industry. The external contexts bring into play indicators and statistical techniques other than publications, patents, and citations. Using regression analysis, for example, one can examine the importance of knowledge and knowledge spill-over for economic development. The relations can be expected to vary among nations and regions. The field-specificity of changes is emphasized as a major driver of the research agenda. In a knowledge-based economy, institutional arrangements can be considered as support structures for cognitive developments.

**Introduction**

In a paper published posthumously in *Research Policy*, Derek de Solla Price (1984, at p. 6) declared that 'the historiography of "normal" science and of "normal" technology taken together leaves no room for the interaction between science and technology.' For the mediation between these two relatively autonomous developments, Price proposed the concept of 'scientific instrumentalities':

> [B]oth the scientific and the technological innovation may proceed from the same adventitious invention of a new instrumentality. In science the typical result of such a major change is a breakthrough or shift of paradigm. In technology one has a significant innovation and the possibility of products that were not around to be sold last year.

---

[1] Amsterdam School of Communications Research (ASCoR), University of Amsterdam, Kloveniersburgwal 48, 1012 CX Amsterdam, The Netherlands; loet@leydesdorff.net; http://www.leydesdorff.net .
[2] SPRU - Science and Technology Policy Research, University of Sussex, United Kingdom.

As the prime example of an 'instrumentality,' Price elaborated on Galileo's telescope used for the momentous discoveries published in the *Siderius Nuncius* of 1610. However, the author emphasized that an instrumentality—unlike an instrument—could also provide a basis to bind scientists and engineers in invisible colleges through the new methodological and technical options which it makes available. For example, 'statistical techniques such as correlation coefficients, multidimensional scaling, and factor analysis' can be considered as an instrumentality in the social sciences (*Ibid.*, at p. 13).

The organization of these invisible colleges around instrumentalities can be made the subject of systematic reorganization by management in private corporations and/or by S&T policies in the public arena. In a series of studies, Terry Shinn analyzed instrumentation and research technologies as a crucial locus of development in the techno-sciences and as the first candidate for intervention 'between science, state and industry' (Joerges & Shinn, 2001). He argued that these research technologies function as interfaces and thus allow for both integration and differentiation: the community operating at an interface can be maintained historically if its role and function for both sides can generate further innovations. Interface management, thus, is not primarily a "blurring of contexts" (Nowotny *et al*., 2001), but a careful elaboration of functionalities for organizing innovation in relation to institutional constraints (Shinn & Lamy, forthcoming).

Innovations take place at interfaces. Thus, a non-linear dynamics is involved which shapes new layers as the outcome of interactions and stabilizes them recursively over time if the institutional conditions can be constructed as support structures (Bathelt, 2003; Cooke & Leydesdorff, 2006). The innovative construction of institutions became legitimate only after the French Revolution, when the organization of the state apparatus was secularized. The latter half of the $19^{th}$ century witnessed the 'wedding of the sciences and the useful arts,' for example, in industrial R&D laboratories (Noble, 1977), in parallel with the 'scientific-technical revolution' of managerial capitalism at the macro level. Braverman (1974) described the systemic character of these processes of change as follows:

> The scientific-technical revolution, for this reason, cannot be understood in terms of specific innovations—as in the case of the Industrial Revolution, which may be adequately characterized by a handful of key inventions—but must be understood rather in its totality as a mode of production into which science and exhaustive engineering investigations have been integrated as part of ordinary functioning. The key innovation is not to be found in chemistry, electronics, automatic machinery, aeronautics, atomic physics, or any of the products of these science-technologies, but rather in the transformation of science into capital. (*Ibid*., pp. 166f.)

These transformations in the late 19th century followed upon the development of a system of nation-states in Europe, Japan, and the U.S.A. (Leydesdorff, 1997). While markets can be relatively protected by state boundaries, science-based innovation could become a driver of competition on the world market (Schumpeter, 1912). For example, the First World War made it possible to boost the American chemical industry by means of the seizure and redistribution of German patents (Noble, 1977).

**Interfaces between different sub-dynamics**

Because science and technology develop at the macro level, the organization of interfaces between science and industry during the 20th century appealed increasingly for state intervention. Particularly after the Sputnik shock of 1957, the OECD was transformed into a supra-governmental instrument to stimulate S&T policies (Freeman, 1982). The European Union made innovation policies the core of its strategic ambition to develop a knowledge-based economy (EC, 2000; Foray & Lundvall, 1996; Leydesdorff, 2005). However, science, technology, and industry involve different logics, and therefore pressures for reorganization are continuously generated at the interfaces between these developments (Dits & Berkhout, 1999). Reorganizations can be smoothened informally, that is, through social relations among industrialists, scientists, and engineers, or regulated formally, for example, through patent legislation (Granstrand, 1999; Jaffe & Trajtenberg, 2002; Van den Belt & Rip, 1987) and, after the Second World War, by means of institutionalized S&T policies (Rothwell & Zegveld, 1981).

In summary, three subdynamics have to be secured in an advanced innovation system: (1) wealth generation in economic exchange relations (Keynes, 1936); (2) novelty production upsetting the equilibrium-seeking mechanisms of the market (Schumpeter, 1943; Nelson & Winter, 1982); and (3) the organization of the social system at the public/private interface (Freeman & Perez, 1988). The emergence of S&T policies in the post-war decades gave the state an increasing role in the development of systems of innovation. The Triple Helix model acknowledges this role of governments without presuming that innovation systems should therefore be considered *ex ante* as 'national' (Etzkowitz & Leydesdorff, 2000; cf. Lundvall, 1992; Nelson, 1993).

The three interacting subdynamics can be expected to contain structures which function as selection environments for one another (McKelvey, 1997). First, the market operates as a clearing mechanism searching for equilibrium between supply and demand at each moment of time. Secondly, innovations operate over time and can be based on longer-term developments in science and technology, which are organized into disciplines and industrial sectors, respectively (Whitley, 1984; Pavitt, 1984). Thirdly, an institutional apparatus has to be developed at the level of (transnational) corporations and the state (Galbraith, 1967). However, the three subdynamics are reflected using different metrics: (1) econometrics, (2) scientometrics and patent statistics, and (3) national (e.g., labor and production) statistics. The instrumentalities involved in these three traditions (e.g., differential equations, descriptive statistics, and regression analysis) provide common ground for the understanding, but their formalization cannot provide sufficient guidance for the measurement of innovations at interfaces, innovation systems, or a knowledge-based economy (Carter, 1996; OECD/Eurostat, 1997). From their different perspectives in the relevant disciplines, researchers have specified other units of analysis (Leydesdorff *et al.*, 1996).

For example, evolutionary and institutional economists have focused on firms and entrepreneurship (e.g., Nelson & Winter, 1982; Casson, 1997), while scientometricians are interested in the development of research fronts at the macro level, using patents, publications, and citations as indicators of contributions by these (and other) institutional units. Following Freeman's (1986) study of Japan, Lundvall (1988) proposed that *national* systems be considered as a first candidate for such analysis. However, he formulated his claim of 'national systems of innovation' pragmatically in terms of a heuristics:

> The interdependency between production and innovation goes both ways. […] This interdependency between production and innovation makes it legitimate to take the national system of production as a starting point when defining a system of innovation. (*Ibid.*, at p. 362)

The choice for the nation as unit of analysis enabled Lundvall to integrate evolutionary economics with institutional and industrial economics (e.g., the analysis of 'filières'; De Bandt & Humbert, 1985), but he also noted that despite their mutual interdependence, national systems of production 'differ in important respects' from innovation processes (p. 362). However, he failed to specify this difference in terms of an analysis of socio-cognitive developments as a selection environment with dynamics different from those of nation-states or global markets (Lundvall & Boras, 1997; Leydesdorff & Meyer, 2003).

Within the research tradition of evolutionary economics, Andersen (1994) noted that from an evolutionary perspective, one should first raise the question of 'what is the evolving system?' Social coordination mechanisms like markets and sciences can evolve, but nations and institutions mainly provide a framework for stabilization, that is, they function as retention mechanisms. Institutions and organization can also be considered as quasi-equilibria which provide buffers against market fluctuations (Aoki, 2001). They thus add another equilibrating sub-dynamics to counter-act the upsetting dynamics of knowledge-based innovations.

The interference of organizations in markets is accordingly accounted in the economic model as *transaction costs* (Williamson, 1985). Costs for R&D, however, can be considered as *investment costs* (Freeman & Perez, 1988). Cowan & Foray (1997) argued that the construction of a knowledge base is achieved by sustained investments in codification processes (Foray, 2004). The measurement of these processes in terms of their information dynamics has been the primary focus of scientometrics.

**Codification processes in knowledge-based developments**

Codification processes can be expected to have effects both within organizations—and thus interact with transaction costs, for example, in the case of process innovations—and at the level of the market, for example, in terms of generating new products. Figure 1 elaborates the relevant processes in the knowledge-production domain using the three dimensions of positions, exchanges, and reflections which we have used in previous visualizations of the Triple Helix dynamics (Leydesdorff, 1990, 1995, 1998; Leydesdorff & Meyer, 2003).

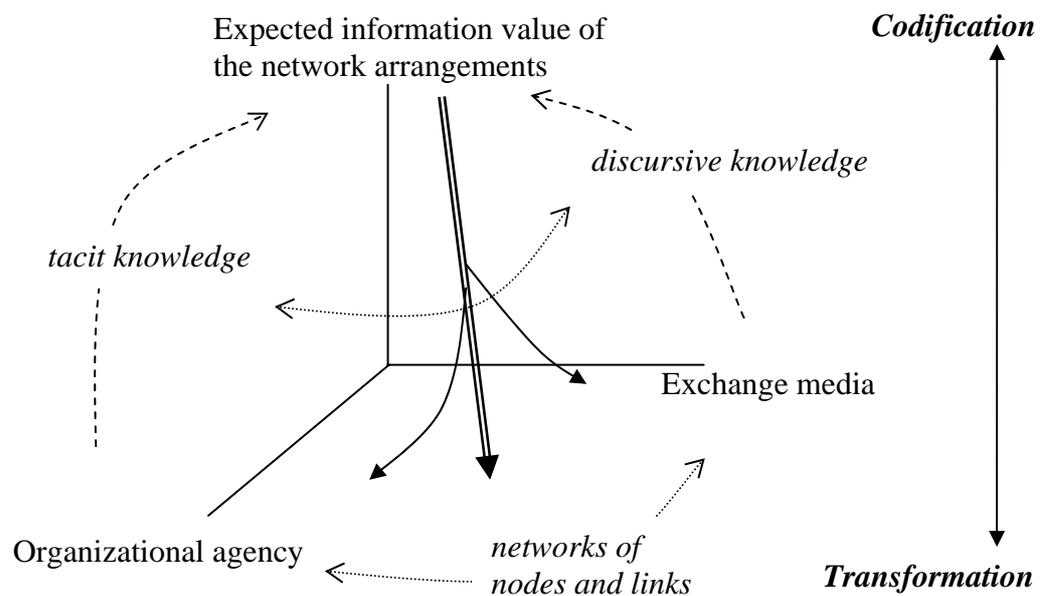

**Figure 1**: The non-linear dynamics of codification and knowledge-based transformation processes

Organizations and agents learn during exchanges and thus develop tacit knowledge. Discursive knowledge can be considered as a medium of exchange different from that of economic exchanges (Luhmann, 1989, 1990): while the market equilibrates in terms of prices at each moment of time, scientific expectations can be stabilized over time. Thus, the interactions between nodes and links in the resulting networks of knowledge production and control stand (analytically) perpendicular to the interactions in an economy.

Knowledge-based interactions can transform the economic ones if another (third) interface between these two types of interactions can be organized and institutionalized. The organization of such interfaces at the level of nation-states and (transnational) corporations can be expected to transform both the economy and its knowledge-base. When interaction effects interact, a non-linear dynamics is generated (Leydesdorff & Meyer, 2003 and forthcoming). Inputs no longer relate to outputs when non-linear dynamics prevail.

**Evaluation of different subdynamics**

How can this complex system—complex because composed of several and interacting sub-dynamics—be evaluated? The historical progression towards a knowledge-based economy varies among nations; integration at the national level still plays a major role in any system of innovation (Skolnikoff, 1993; Riba-Vilanova and Leydesdorff, 2001). However, transnational levels of government like the European Union and the ongoing devolution of nations into regions have changed the functions of national governments (Braczyk *et al*., 1998; Cooke & Leydesdorff, 2006). While national governments were previously integrators in institutional terms, 'governance' nowadays spans a variety of sub- and supranational levels (Kooiman, 1993). However, these polycentric environments remain highly structured: puzzles have continuously to be solved at interfaces by innovation, and the interfacing systems themselves remain under pressure to change (Larédo, 2003).

Innovations can be considered as variations that have been stabilized under the selection pressure of these different environments. Thus, innovations can be considered as performative integrations. A new medication, for example, has to be warranted by its science base, patent-protected and government-approved, and also commercially viable. All these aspects require specific expertise, and these operations can be evaluated in terms of system-specific statistics. The selection environments are themselves dynamic. However, they can be expected to change at a slower pace. New science-based developments, for example, may require changes in the rules and regulations at the structural level. For example, countries other than the U.S.A. felt it necessary to pass new patent legislation after the Bayh-Dole Act of 1980. Structural change, however, can be expected to operate at a frequency lower than that of operational change. Structures select deterministically on variation at each moment of time. While the selection mechanisms operate in a distributed mode, these deterministic mechanisms remain probabilistic.

In summary, an internal differentiation and integration of science and technology focusing on instrumentalities can be distinguished from the external differentiation and integration in the relations of this knowledge production process to other social domains, notably governance and industry. These external contexts bring other indicators and statistical techniques into play, such as regression analysis. For example, one can ask how important is knowledge or knowledge spill-over for economic development? This relationship can be expected to vary among nations and regions.

**Geographic units of analysis**

The studies assembled in this special issue were first presented at the Fifth International Conference on the Triple Helix in Turin, May 2005. In the Triple Helix context, institutional units of analysis prevail because of the focus on evaluation. For example, *Poh-Kam Wong and Yuen-Ping Ho,* in their study entitled 'Knowledge Sources of Innovation in a Small Open Economy: The Case of Singapore,' adopt the model of a national system of innovation in order to trace the sources and flows of codified knowledge using citation data of patents and publications. How can the portfolios of multinational corporations, international science, and national capacities be integrated and recombined in the context of Singapore as a hub? What is the role of indigenous knowledge in this process?

At a lower level of aggregation, one can compare regions in terms of patenting and publishing. In their paper entitled 'In which regions do universities patent and publish more?' *Joaquín Azagra-Caro, Fragiskos Archontakis, and Alfredo Yegros-Yegros* build university production functions for 17 Spanish regions and use an econometric model to estimate their determinants. They reach the interesting conclusion that university patenting follows R&D expenditure, while the number of publications is dependent on the number of researchers. The latter factor is long-term and thus less sensitive to government intervention, but the former can be facilitated in the short term. Investment policies in the knowledge base of an economy, therefore, require a trade-off between these two perspectives.

In a short communication, *Wolfgang Glänzel and Balász Schlemmer* compare national research profiles for six small European countries, of which three (Ireland, Finland, and Portugal) belonged to the EU-15, and three (Hungary, Estonia, and Slovenia) were accession countries in 2005. The data were carefully assigned in terms of institutional addresses, and Triple Helix relations were subsequently measured in terms of co-authorship relations among the sectors. The results show huge differences in the *ex ante* situation of 1983, which are more moderate during the next period (1993 and 2003 were used as other points for the measurement). As could be expected, the patterns of development in Western European countries are more continuous than in the accession countries. For example, in the EU-15 member states the shares of publications with exclusively university addresses has been approximately 80%, while in 2003 this percentage remained much lower for Hungary (52.4%) and Slovenia (55%). Have policy inputs to stimulate university-industry collaborations been more effective in the transition economies than in Western Europe?

**The emergence of the entrepreneurial university**

A number of studies assume national contexts for the sample choice, but the case studies propose to generalize their findings about collaboration. Is an 'entrepreneurial university' emerging? The entrepreneurial university (Clark, 1998; Etzkowitz, 2002) is one which extends its missions in higher education and academic research to assume the role of stimulating economic innovation in the environment. A new role pattern for academicians would thus be required.

In their paper entitled 'Industrial Linkages in Indian Universities: What they reveal and what they imply?' *Sujit Bhattacharya and Praveen Arora* examine motivating factors for collaboration with industry among university departments in seven Indian universities. The study concludes that personal contacts are indicated as the major motivation in the initiation of linkages, but that special centers facilitate the transfer. In a survey of 1,554 Canadian university scholars who received funding by the Natural Sciences and Engineering Research Council of Canada, *Omar Belkhodja and Réjean Landry* conclude that the factors explaining the decision to collaborate with industry and government vary with the fields of study. A set of factors can be distinguished (e.g., the researcher's strategic position, the network structures, and the costs involved) which may function as facilitators or impediments.

Using a sample of 208 Italian professors who were involved in patenting, *Nicola Baldini, Rosa Grimaldi, and Maurizio Sobrero* conclude that the cognitive goals of the research process and prestige are the main drivers for patenting, and not the economic earnings. Because of the many obstacles encountered by academics asking to obtain patents, regulation at the level of the university is considered helpful. A final study in this set is provided by *Paula Susana Figueiredo Moutinho, Margarida Fontes, and Manuel Mira Godinho,* who questioned Portuguese scientists in public-sector research institutes about their attitudes towards patenting. The majority of these researchers has no 'ethical' objections against patenting, but expected only weak personal or professional benefits from it. Most of the respondents expected huge difficulties because of the lack of institutional support for patenting.

In summary, these studies conclude that patenting is not a core task of academics, but it can be used insofar as it serves to further research and accordingly adds to a scientist's prestige and reputation. There is no longer resistance to this additional channel of communication, but researchers have a realistic estimate about the amount of additional effort which patenting may require.

**Academic Patenting**

Although patenting thus seems to have remained relatively marginal as an asset within the reputationally controlled reward structure of science (Whitley, 1984), the rise in university patenting over the last decades has been structural and therefore a subject of study. Inventor-author relations are not easily retrievable and one may underestimate relations limited to name searches. In their study entitled 'Measuring Industry-Science Links through Inventor-Author relations,' *Bruno Cassiman, Patrick Genisson, and Bart Van Looy* develop a text-based profiling methodology that improves both the recall and the precision of linkages. *Eric Iversen, Magnus Gulbrandsen, and Antje Klitkou* propose a three-stage procedure which should provide a baseline for the measurement of the impact of academic patenting legislation in Norway.

In an evaluation of how patents are valued within academia, *Martin Meyer and Puay Tang* conclude that no sufficient standard of measurement has yet been achieved for developing a robust indicator. In their paper entitled 'Developing Technology in the Vicinity of Science,' *Bart van Looy, Tom Magerman, and Koenraad Debackere* argue (in the case of biotechnology) that there is a mutual positive relationship between scientific and technological productivity. Technological productivity is associated with the science-intensity of patents.

In summary, in accordance with the subsidiary status of patents within academia, the relationship between university research and patents has yet to be fully understood. Practices vary among countries and disciplines. Verspagen (2006) found that in a substantial percentage of cases, researchers at European universities did not patent, but left the intellectual property to their industrial partners. Even in countries where universities hold the commercial right to faculty inventions, more than half of the academic patents are still assigned exclusively to industry. For example, Meyer *et al*.

(2005) report that approximately 58% of the inventions at Flemish universities which resulted in U.S. patents between 1996 and 2001, were owned by corporations. One can expect that this percentage is even higher in national systems in which not the universities, but academic faculty holds the patent rights because of the problems indicated in the various contributions. Accordingly, the field of determining the knowledge base of patents is still very much in development (Leydesdorff, 2004).

**Science-technologies as units of analysis**

The final part of the issue collects three studies which focus on a specific technology. In a study entitled 'Networks of Knowledge: The Distributed Nature of Medical Innovation,' *Ronnie Ramlogan, Andrea Mina, Gindo Tampubolon, and Stan Metcalfe* map the trajectories of research in two clinical areas ('Coronary Artery Disease' and 'Glaucoma') in terms of epistemic, geographical, and organizational distributions. According to these authors, rich ecologies which facilitate collaborations within and among institutions are crucial for medical innovation. The question of setting boundaries arises in a policy context, for example, because of considerations about cost control.

Focusing on Norway as their empirical domain, *Antje Klitkou, Stian Nygaard, and Martin Meyer* raise the question of 'tracking techno-science networks' in the case of fuel cells and hydrogen-technology related R&D. Most authors in the sample are active both in science and technology; their research activities are internationally organized and woven into networks of co-authorship and co-patenting. From this perspective, a national innovation system can be considered as a specific density— among other possible densities—in a multi-dimensional and multi-layered network. The network is driven by substantive sources of variance (e.g., expectations of success) which motivate scholars to cross institutional and disciplinary boundaries. Evaluation from the perspective of a national system might therefore lose its meaning and even become counterproductive. Evaluation has to focus on output and not on institutional parameters.

The Internet has reinforced the capacity of scientists and engineers to network beyond institutional and other (e.g., intellectual) boundaries. Knowledge flows and knowledge spill-over may be less containable within organizations that are not organized in terms of the structures of the knowledge production process like specialties and disciplines. How might these processes—self-organizing across institutional boundaries—nevertheless be mapped? *Hildrun, Ute, and Theo Kretschmer* in a study entitled 'Reflection on Co-authorship Networks in the Web,' argue that hyperlinks, while formally analogous to citations, cannot indicate intellectual lineages and collaborations because of institutional distortion. Coauthored publications, however, can be used as seeds for web citations. A set of 'Web Visibility Indicators' is proposed.

**A scientometric map of this Triple Helix issue**

In addition to the above reflections on the various contributions to this special issue, we can also use the tools of scientometrics itself to analyze whether and how the Triple Helix is a loosely linked epistemic community of researchers from various disciplines and specialties of the social sciences (and beyond) who share an interest in knowledge exchange processes between university, industry, and government. Can the Triple Helix be considered as a metaphor that creates a platform for discourse and integrates researchers from quite different backgrounds, acting as 'interpersonal stabilizer', as Marz and Dierkes (1994) put it? Or is this perhaps a mixed bag which we generated artificially without much internal cohesion? (Leydesdorff & Hellsten, 2005).

Previous studies exploring the Triple Helix community in bibliometric terms (e.g. Shinn, 2000; Glennisson *et al*. 2005) compared contributions with reference to related concepts like 'Mode 2' and national innovation systems (cf. Etzkowitz & Leydesdorff, 2000). While this is not the place to map out the cognitive structure of this entire field in detail, exploring the inter-relations of the contributions to this topical issue may offer some new ideas as to how the community of scientometricians participates in the interdisciplinary discourse about the Triple Helix. It goes without saying that what follows is a modest beginning that requires a more comprehensive and thorough follow-up in the future. In a way we follow here Braun's (2006) example by providing a 'bibliometric review' of this special issue in order to gain more insights in its structure.

To this purpose, bibliographic coupling was applied to the data set of fifteen papers (including this introduction) on the basis of the references which are shared among them. The combined reference sets of the papers in this issue encompass 609 items, which means that on average each paper contains slightly more than 40 references.

After editing the reference set in order to ensure the replacement of variants by a standard, 59 pairs of papers sharing references were identified using BibExcel.[3] These couples were ranked in declining order by the number of references shared. (One pair shared 13 references while 25 pairs shared only one.) Based on this, a (similarity) matrix was created consisting of shared references among each pair of papers. This was used as input to a mapping procedure using ALSCAL for the multi-dimensional scaling. In addition to this, cluster analysis was carried out using Persson's (1994) clustering routine.

Figure 2 provides the resulting map. Thickness of the links between papers indicates the frequency of shared references between them. As one can see from the map, all papers are related and most of the papers share a common platform of references.

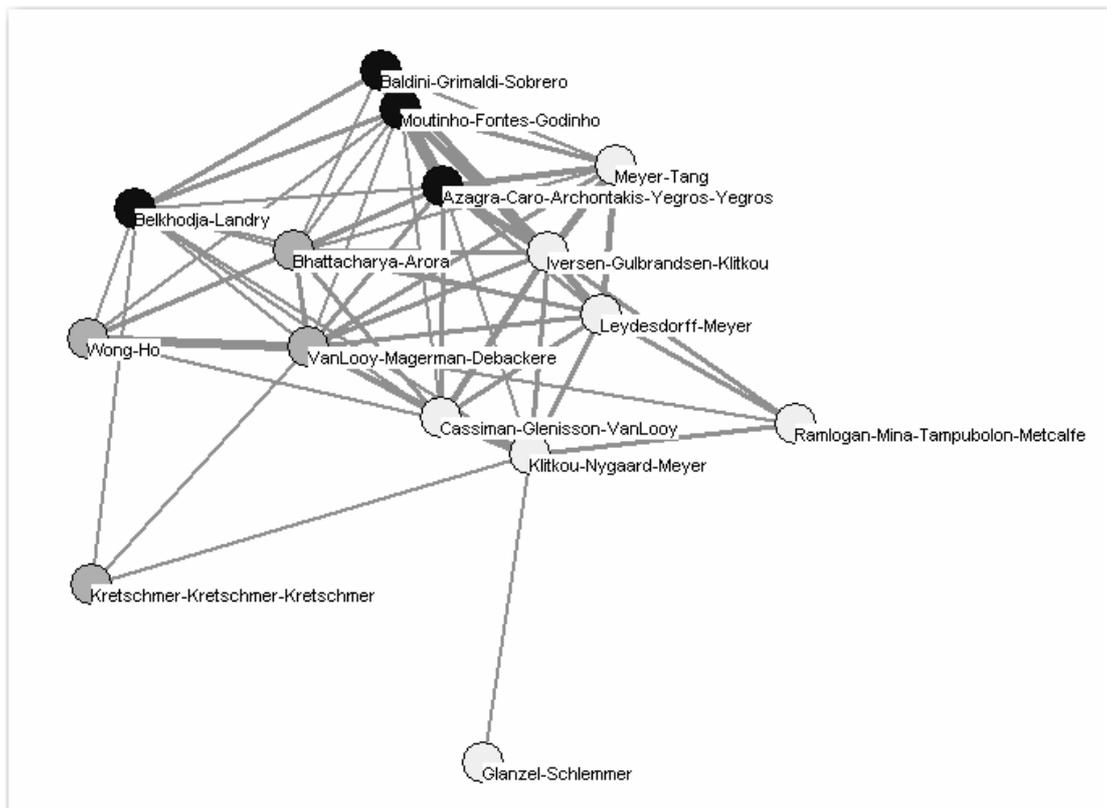

**Figure 2**: Bibliographic coupling map of the 15 contributions to this topical issue

---

[3] BibExcel is freeware for academic purposes, available at http://www.umu.se/inforsk/Bibexcel/

Using the clustering algorithm of BibExcel, three clusters are distinguished: a first cluster comprises the papers of *Baldini et al.*, *Moutinho et al.*, *Azagra-Caro et al*, extending to *Belkhodja and Landry*. Another cluster is formed by these papers: *Iversen et al.*, *Leydesdorff & Meyer*, *Meyer & Tang*, *Ramlogan et al.*, *Cassiman et al.*, *Klitkou et al.*, and *Glänzel & Schlemmer*. The third cluster comprises three papers: *Van Looy et al.*, *Wong & Ho*, and *Bhattacharya & Arora*.

Arguably, the reference-based links between these papers reflect shared interests: The papers in the first cluster focus on patenting in universities and other public research organizations. At the core of this cluster, there is a strong link between the papers by *Baldini et al.* and *Moutinho et al.* as well as *Azagra-Caro et al.* These papers analyze academic patenting and the attitudes of researchers in a European context. Both papers share with the third a strong appreciation of economics and econometrics-oriented contributions to this broader area of research. The common reception of (mostly quantitative) studies of university-industry collaboration is what links these papers to the work by *Belkhodja & Landry*.

The papers indicated as the second, largest and somewhat more diverse cluster share a common interest in approaches to track links between science and technology, boundary-crossing networks, and a stronger appreciation or discussion of approaches associated with 'systems of innovation' and the 'new production of knowledge'. The methodological link is especially strong between *Cassiman et al.*, *Klitkou et al.*, and to some extent also *Iversen et al*. These three papers share an interest in linking science and technology through tracking co-active researchers. This introduction is inter-linked with other papers in this cluster through addressing these issues from similar theoretical perspectives. In this cluster, papers by authors with an evolutionary economics background appear to play a more prominent role than in the first cluster. Not surprisingly, co-evolving networks in science and technology is another linking theme in this cluster. This concerns especially the papers by *Klitkou et al.* and *Ramlogan et al.*

The third and final cluster brings together papers that share an interest in patent citation analysis as a way of linking science and technology (in particular, *Van Looy et al.* and *Wong & Ho*). Patent citations as indicators of regional knowledge spillovers link these papers to *Bhattacharya & Arora*'s contribution.

This small bibliometric exercise suggests that the Triple Helix provides a field or community crossing boundaries and providing interfaces. Drawing on notions introduced earlier in this introduction, 'differentiation' (signified by the clusters in this map) seems to coincide with 'integration' (as traced in the manifold of links among the papers in different clusters). It is difficult to generalize from these observations, but perhaps this effort provides some context or insights for future discussions about a reflexive contribution of scientometrics to the Triple Helix discourse.

**Conclusions**

The function of organized knowledge production and control systems for the economy and society at large has changed structurally during the last two decades. After the oil crises of the 1970s, advanced industrial economies became dependent increasingly on knowledge as a source of innovation. The sciences penetrated other social domains no longer only as a source of innovation using a linear model, but as non-linear effects of interactions at interfaces with other social domains.

The focus on innovation changed the position of universities. This was first reflected in the U.S.A., for example, with the introduction of the Bayh-Dole Act in 1980 allowing universities to apply for patents on the basis of federal funding. Other countries had to follow suit by rethinking their research portfolio, institutional make-up, and legislation about intellectual property rights. The European Union forcefully made 'innovation' the topic of its consecutive Framework Programs during the 1980s and 1990s, while traditional science policies were left to elite institutions at the national level (e.g., research councils). In this context, concepts like 'Mode 2' and the Triple Helix could function as a wake-up call during the 1990s.

In the meantime, the dust has settled. University-industry relations have now been accepted, and patenting by universities has reached a stable level (Leydesdorff & Meyer, forthcoming). Transfer offices have been brought into place. The creation of a knowledge-based economy has become an accepted objective of government policies around the globe. In Leydesdorff & Meyer (2003) we submitted the Triple Helix model as an analytical tool for the study of these complex dynamics. In this issue and introduction, we focus more than in the previous one on the specificity of codification along the cognitive axis. Institutional arrangements can be expected to follow cognitive leads such as instrumentalities at interfaces because the evolution of the knowledge-based system is driven by options for innovation. For example, the current wave of nanotechnology can be expected to change university-industry-government relations. The mapping and visualization of these changes remains a task for the information sciences. The field-specificity of the changes is emphasized in various contributions as a major driver of the research agenda.

## Acknowledgements

We thank Terry Shinn and Olle Persson for comments on previous versions of this manuscript. We are grateful to Jenny Newton for her editorial assistance.